\documentclass[conference]{IEEEtran}
\usepackage{latexsym,epsf,times,amsmath,color,amssymb,indentfirst,subfigure,fancyhdr,colortbl,bm,caption}
\usepackage{graphicx}
\usepackage{algorithm}
\usepackage{algorithmic}
\usepackage{psfrag}
\newcommand{\nc}{\newcommand}
\nc{\be}{\begin{equation}}
\nc{\ee}{\end{equation}}
\nc{\beqa}{\begin{eqnarray}}
\nc{\eeqa}{\end{eqnarray}}
\def\calA{{\mathcal{A}}}

\def\calQ{{\mathcal{K}}}

\nc{\cv}{{\mathbf{c}}}
\nc{\cbf}{\boldsymbol{c}}
\nc{\mubf}{\boldsymbol{\mu}}
\nc{\kESM}{$\kappa$ESM}
\nc{\gammabf}{{\mbox{\boldmath$\gamma$}}}
\nc{\varphibf}{{\mbox{\boldmath$\varphi$}}}
\nc{\varthetabf}{{\mbox{\boldmath$\vartheta$}}}
\nc{\lambdabf}{{\mbox{\boldmath$\lambda$}}}
\nc{\Thetabf}{{\mbox{\boldmath$\Theta$}}}
\nc{\etabf}{{\mbox{\boldmath$\eta$}}}
\nc{\taubf}{{\mbox{\boldmath $\tau$}}}
\nc{\alphabf}{{\mbox{\boldmath $\alpha$}}}
\nc{\nubf}{{\mbox{\boldmath$\nu$}}}
\nc{\xibf}{{\mbox{\boldmath$\xi$}}}
\nc{\rhobf}{{\mbox{\boldmath$\rho$}}}
\nc{\Lambdabf}{{\mbox{\boldmath$\Lambda$}}}
\nc{\Gammabf}{{\mbox{\boldmath$\Gamma$}}}
\nc{\Upsilonbf}{{\mbox{\boldmath$\Upsilon$}}}
\nc{\pbf}{{\mbox{\boldmath $p$}}}
\nc{\mbf}{{\mbox{\boldmath $m$}}}
\nc{\zerobf}{{\mbox{\boldmath $0$}}}
\nc{\fbf}{{\mbox{\boldmath $f$}}}
\nc{\sbf}{{\mbox{\boldmath $s$}}}
\nc{\xbf}{{\mbox{\boldmath $x$}}}
\nc{\zbf}{{\mbox{\boldmath $z$}}}
\nc{\ybf}{{\mbox{\boldmath $y$}}}
\nc{\bbf}{{\mbox{\boldmath $b$}}}
\nc{\uvbf}{{\mbox{\boldmath $1$}}}
\nc{\wbf}{{\mbox{\boldmath $w$}}}
\nc{\vbf}{{\mbox{\boldmath $v$}}}
\nc{\gbf}{{\mbox{\boldmath $g$}}}
\nc{\abf}{{\mbox{\boldmath $a$}}}
\nc{\wbfs}{{\mbox{\boldmath \scriptsize $w$}}}
\nc{\Hbf}{{\mbox{\boldmath$H$}}}
\nc{\Abf}{{\mbox{\boldmath$A$}}}
\nc{\Gbf}{{\mbox{\boldmath$G$}}}
\nc{\Rbf}{{\mbox{\boldmath$R$}}}
\nc{\Ubf}{{\mbox{\boldmath$U$}}}
\nc{\Vbf}{{\mbox{\boldmath$V$}}}
\nc{\Ibf}{{\mbox{\boldmath$I$}}}
\nc{\diag}{\text{diag}}
\nc{\Diag}{\text{Diag}}

\usepackage{amssymb, amsmath}
\usepackage{amsfonts}
\usepackage{bbm}
\usepackage{mathrsfs}
\usepackage{xspace}
\usepackage{bm}

\usepackage{cite}
\usepackage{epsfig}
\usepackage{cases}
\usepackage{psfrag}
\usepackage{enumerate}

\usepackage{makeidx}
\usepackage{supertabular}

\newenvironment{textbmatrix}{   \setlength{\arraycolsep}{2.5pt}%
                                                                \big[\begin{matrix}}{\end{matrix}\big]%
                                                                \raisebox{0.08ex}{\vphantom{M}}}

\def\be{\begin{equation}}
\def\ee{\end{equation}}
\def\een{\nonumber \end{equation}}
\def\mat{\begin{bmatrix}}
\def\emat{\end{bmatrix}}
\def\btm{\begin{textbmatrix}}
\def\etm{\end{textbmatrix}}

\def\ba#1\ea{\begin{align}#1\end{align}}
\def\bs#1\es{\begin{split}#1\end{split}} 
\def\bg#1\eg{\begin{gather}#1\end{gather}} 
\def\bi#1\ei{\begin{itemize}#1\end{itemize}}

\newcommand{\safemath}[2]{\newcommand{#1}{\ensuremath{#2}\xspace}}

\safemath{\interior}{\mathrm{Int}}                       % interior of a set
\safemath{\dfn}{:=}                                                     % definition
\safemath{\dirac}{\delta}                                       % Dirac delta

\safemath{\SNR}{\text{\sc snr}}                                 % signal to noise ratio
\safemath{\No}{N_0}                                                     % noise spectral density
\safemath{\Es}{E_s}                                                     % energy per symbol
\safemath{\Eb}{E_b}                                                     % energy per bit
\safemath{\EbNo}{\frac{\Eb}{\No}}
\safemath{\EsNo}{\frac{\Es}{\No}}

\DeclareMathOperator{\CHop}{\ensuremath{\mathbb{H}}} % channel operator
\safemath{\tvir}{h_{\CHop}}                                     % time-varying impulse response
\safemath{\tvtf}{L_{\CHop}}                                     %       transfer function
\safemath{\spf}{S_{\CHop}}                                              
\safemath{\bff}{H_{\CHop}}                                      %       function

\safemath{\ircf}{R_{h}}                                         % impulse response correlation fn.
\safemath{\scf}{R_{S}}                                          % scattering function
\safemath{\tfcf}{R_{L}}                                         % time-frequency correlation fn.
\safemath{\bfcf}{R_{H}}                                         % bi-frequency correlation fn.
\safemath{\mi}{I}                                                       % muttal information
\safemath{\capacity}{C}                                         % capacity

\newcommand{\pdf[2]}{f_{{#1}}\left({#2}\right)}                    % probability density function
\safemath{\uniform}{\mathcal{U}}                        % uniform distribution
\safemath{\normal}{\mathcal{N}}                         % normal distribution
\safemath{\circnorm}{\mathcal{CN}}                      % circ. symm. normal
\safemath{\mchain}{\leftrightarrow}                     % Markov chain

\safemath{\dB}{\,\mathrm{dB}}
\safemath{\dBm}{\,\mathrm{dBm}}
\safemath{\Hz}{\,\mathrm{Hz}}
\safemath{\kHz}{\,\mathrm{kHz}}
\safemath{\MHz}{\,\mathrm{MHz}}
\safemath{\GHz}{\,\mathrm{GHz}}
\safemath{\s}{\,\mathrm{s}}
\safemath{\ms}{\,\mathrm{ms}}
\safemath{\mus}{\,\mathrm{\mu s}}
\safemath{\ns}{\,\mathrm{ns}}
\safemath{\meter}{\,\mathrm{m}}
\safemath{\km}{\,\mathrm{km}}
\safemath{\mm}{\,\mathrm{mm}}
\safemath{\cm}{\,\mathrm{cm}}
\safemath{\m}{\,\mathrm{m}}
\safemath{\W}{\,\mathrm{W}}
\safemath{\J}{\,\mathrm{J}}
\safemath{\K}{\,\mathrm{K}}
\safemath{\bit}{\,\mathrm{bit}}
\safemath{\nW}{\,\mathrm{nW}}
\safemath{\muW}{\,\mathrm{$\mu$W}}
\safemath{\Watt}{\,\mathrm{W}}

\safemath{\define}{\triangleq}                  % definition

\safemath{\equivalent}{\sim}
\safemath{\distas}{\sim}                                        % distributed according to

\safemath{\reals}{\mathbb{R}}
\safemath{\positivereals}{\mathbb{R}^{+}}
\safemath{\integers}{\mathbb{Z}}
\safemath{\posint}{\mathbb{Z}_{+}}
\safemath{\naturals}{\mathbb{N}}
\safemath{\complexset}{\mathbb{C}}

\safemath{\setA}{\mathcal{A}}
\safemath{\setB}{\mathcal{B}}
\safemath{\setC}{\mathcal{C}}
\safemath{\setD}{\mathcal{D}}
\safemath{\setE}{\mathcal{E}}
\safemath{\setF}{\mathcal{F}}
\safemath{\setG}{\mathcal{G}}
\safemath{\setH}{\mathcal{H}}
\safemath{\setI}{\mathcal{I}}
\safemath{\setJ}{\mathcal{J}}
\safemath{\setK}{\mathcal{K}}
\safemath{\setL}{\mathcal{L}}
\safemath{\setM}{\mathcal{M}}
\safemath{\setN}{\mathcal{N}}
\safemath{\setO}{\mathcal{O}}
\safemath{\setP}{\mathcal{P}}
\safemath{\setQ}{\mathcal{Q}}
\safemath{\setR}{\mathcal{R}}
\safemath{\setS}{\mathcal{S}}
\safemath{\setT}{\mathcal{T}}
\safemath{\setU}{\mathcal{U}}
\safemath{\setV}{\mathcal{V}}
\safemath{\setW}{\mathcal{W}}
\safemath{\setX}{\mathcal{X}}
\safemath{\setY}{\mathcal{Y}}
\safemath{\setZ}{\mathcal{Z}}
\safemath{\emptySet}{\varnothing}

\safemath{\bma}{\mathbf{a}}
\safemath{\bmb}{\mathbf{b}}
\safemath{\bmc}{\mathbf{c}}
\safemath{\bmd}{\mathbf{d}}
\safemath{\bme}{\mathbf{e}}
\safemath{\bmf}{\mathbf{f}}
\safemath{\bmg}{\mathbf{g}}
\safemath{\bmh}{\mathbf{h}}
\safemath{\bmi}{\mathbf{i}}
\safemath{\bmj}{\mathbf{j}}
\safemath{\bmk}{\mathbf{k}}
\safemath{\bml}{\mathbf{l}}
\safemath{\bmm}{\mathbf{m}}
\safemath{\bmn}{\mathbf{n}}
\safemath{\bmo}{\mathbf{o}}
\safemath{\bmp}{\mathbf{p}}
\safemath{\bmq}{\mathbf{q}}
\safemath{\bmr}{\mathbf{r}}
\safemath{\bms}{\mathbf{s}}
\safemath{\bmt}{\mathbf{t}}
\safemath{\bmu}{\mathbf{u}}
\safemath{\bmv}{\mathbf{v}}
\safemath{\bmw}{\mathbf{w}}
\safemath{\bmx}{\mathbf{x}}
\safemath{\bmy}{\mathbf{y}}
\safemath{\bmz}{\mathbf{z}}

\bmdefine{\biad}{a}
\bmdefine{\bibd}{b}
\bmdefine{\bicd}{c}
\bmdefine{\bidd}{d}
\bmdefine{\bied}{e}
\bmdefine{\bifd}{f}
\bmdefine{\bigd}{g}
\bmdefine{\bihd}{h}
\bmdefine{\biid}{i}
\bmdefine{\bijd}{j}
\bmdefine{\bikd}{k}
\bmdefine{\bild}{l}
\bmdefine{\bimd}{m}
\bmdefine{\bind}{n}
\bmdefine{\biod}{o}
\bmdefine{\bipd}{p}
\bmdefine{\biqd}{q}
\bmdefine{\bird}{r}
\bmdefine{\bisd}{s}
\bmdefine{\bitd}{t}
\bmdefine{\biud}{u}
\bmdefine{\bivd}{v}
\bmdefine{\biwd}{w}
\bmdefine{\bixd}{x}
\bmdefine{\biyd}{y}
\bmdefine{\bizd}{z}

\bmdefine{\bixid}{\xi}
\bmdefine{\bilambdad}{\lambda}
\bmdefine{\bimud}{\mu}
\bmdefine{\bithetad}{\theta}
\bmdefine{\biphid}{\phi}

\safemath{\bmia}{\biad}
\safemath{\bmib}{\bibd}
\safemath{\bmic}{\bicd}
\safemath{\bmid}{\bidd}
\safemath{\bmie}{\bied}
\safemath{\bmif}{\bifd}
\safemath{\bmig}{\bigd}
\safemath{\bmih}{\bihd}
\safemath{\bmii}{\biid}
\safemath{\bmij}{\bijd}
\safemath{\bmik}{\bikd}
\safemath{\bmil}{\bild}
\safemath{\bmim}{\bimd}
\safemath{\bmin}{\bind}
\safemath{\bmio}{\biod}
\safemath{\bmip}{\bipd}
\safemath{\bmiq}{\biqd}
\safemath{\bmir}{\bird}
\safemath{\bmis}{\bisd}
\safemath{\bmit}{\bitd}
\safemath{\bmiu}{\biud}
\safemath{\bmiv}{\bivd}
\safemath{\bmiw}{\biwd}
\safemath{\bmix}{\bixd}
\safemath{\bmiy}{\biyd}
\safemath{\bmiz}{\bizd}

\safemath{\bmxi}{\bixid}
\safemath{\bmlambda}{\bilambdad}
\safemath{\bmmu}{\bimud}
\safemath{\bmtheta}{\bithetad}
\safemath{\bmphi}{\biphid}

\safemath{\bA}{\mathbf{A}}
\safemath{\bB}{\mathbf{B}}
\safemath{\bC}{\mathbf{C}}
\safemath{\bD}{\mathbf{D}}
\safemath{\bE}{\mathbf{E}}
\safemath{\bF}{\mathbf{F}}
\safemath{\bG}{\mathbf{G}}
\safemath{\bH}{\mathbf{H}}
\safemath{\bI}{\mathbf{I}}
\safemath{\bJ}{\mathbf{J}}
\safemath{\bK}{\mathbf{K}}
\safemath{\bL}{\mathbf{L}}
\safemath{\bM}{\mathbf{M}}
\safemath{\bN}{\mathbf{N}}
\safemath{\bO}{\mathbf{O}}
\safemath{\bP}{\mathbf{P}}
\safemath{\bQ}{\mathbf{Q}}
\safemath{\bR}{\mathbf{R}}
\safemath{\bS}{\mathbf{S}}
\safemath{\bT}{\mathbf{T}}
\safemath{\bU}{\mathbf{U}}
\safemath{\bV}{\mathbf{V}}
\safemath{\bW}{\mathbf{W}}
\safemath{\bX}{\mathbf{X}}
\safemath{\bY}{\mathbf{Y}}
\safemath{\bZ}{\mathbf{Z}}

\safemath{\bZero}{\mathbf{0}}

\bmdefine{\biAd}{A}
\bmdefine{\biBd}{B}
\bmdefine{\biCd}{C}
\bmdefine{\biDd}{D}
\bmdefine{\biEd}{E}
\bmdefine{\biFd}{F}
\bmdefine{\biGd}{G}
\bmdefine{\biHd}{H}
\bmdefine{\biId}{I}
\bmdefine{\biJd}{J}
\bmdefine{\biKd}{K}
\bmdefine{\biLd}{L}
\bmdefine{\biMd}{M}
\bmdefine{\biOd}{N}
\bmdefine{\biPd}{O}
\bmdefine{\biQd}{P}
\bmdefine{\biRd}{R}
\bmdefine{\biSd}{S}
\bmdefine{\biTd}{T}
\bmdefine{\biUd}{U}
\bmdefine{\biVd}{V}
\bmdefine{\biWd}{W}
\bmdefine{\biXd}{X}
\bmdefine{\biYd}{Y}
\bmdefine{\biZd}{Z}

\bmdefine{\biDelta}{\Delta}
\bmdefine{\biLambda}{\Lambda}
\bmdefine{\biPhi}{\Phi}
\bmdefine{\biSigma}{\Sigma}
\bmdefine{\biOmega}{\Omega}
\bmdefine{\biTheta}{\Theta}

\safemath{\bimA}{\biAd}
\safemath{\bimB}{\biBd}
\safemath{\bimC}{\biCd}
\safemath{\bimD}{\biDd}
\safemath{\bimE}{\biEd}
\safemath{\bimF}{\biFd}
\safemath{\bimG}{\biGd}
\safemath{\bimH}{\biHd}
\safemath{\bimI}{\biId}
\safemath{\bimJ}{\biJd}
\safemath{\bimK}{\biKd}
\safemath{\bimL}{\biLd}
\safemath{\bimM}{\biMd}
\safemath{\bimN}{\biNd}
\safemath{\bimO}{\biOd}
\safemath{\bimP}{\biPd}
\safemath{\bimQ}{\biQd}
\safemath{\bimR}{\biRd}
\safemath{\bimS}{\biSd}
\safemath{\bimT}{\biTd}
\safemath{\bimU}{\biUd}
\safemath{\bimV}{\biVd}
\safemath{\bimW}{\biWd}
\safemath{\bimX}{\biXd}
\safemath{\bimY}{\biYd}
\safemath{\bimZ}{\biZd}

\safemath{\bDelta}{\bielta}
\safemath{\bLambda}{\biLambda}
\safemath{\bPhi}{\biPhi}
\safemath{\bSigma}{\biSigma}
\safemath{\bOmega}{\biOmega}
\safemath{\bTheta}{\biTheta}

\safemath{\veca}{\bma}
\safemath{\vecb}{\bmb}
\safemath{\vecc}{\bmc}
\safemath{\vecd}{\bmd}
\safemath{\vece}{\bme}
\safemath{\vecf}{\bmf}
\safemath{\vecg}{\bmg}
\safemath{\vech}{\bmh}
\safemath{\veci}{\bmi}
\safemath{\vecj}{\bmj}
\safemath{\veck}{\bmk}
\safemath{\vecl}{\bml}
\safemath{\vecm}{\bmm}
\safemath{\vecn}{\bmn}
\safemath{\veco}{\bmo}
\safemath{\vecp}{\bmp}
\safemath{\vecq}{\bmq}
\safemath{\vecr}{\bmr}
\safemath{\vecs}{\bms}
\safemath{\vect}{\bmt}
\safemath{\vecu}{\bmu}
\safemath{\vecv}{\bmv}
\safemath{\vecw}{\bmw}
\safemath{\vecx}{\bmx}
\safemath{\vecy}{\bmy}
\safemath{\vecz}{\bmz}
\safemath{\vecZero}{\bZero}

\safemath{\vecxi}{\bmxi}
\safemath{\veclambda}{\bmlambda}
\safemath{\vecmu}{\bmmu}
\safemath{\vectheta}{\bmtheta}
\safemath{\vecphi}{\bmphi}

\safemath{\matA}{\bA}
\safemath{\matB}{\bB}
\safemath{\matC}{\bC}
\safemath{\matD}{\bD}
\safemath{\matE}{\bE}
\safemath{\matF}{\bF}
\safemath{\matG}{\bG}
\safemath{\matH}{\bH}
\safemath{\matI}{\bI}
\safemath{\matJ}{\bJ}
\safemath{\matK}{\bK}
\safemath{\matL}{\bL}
\safemath{\matM}{\bM}
\safemath{\matN}{\bN}
\safemath{\matO}{\bO}
\safemath{\matP}{\bP}
\safemath{\matQ}{\bQ}
\safemath{\matR}{\bR}
\safemath{\matS}{\bS}
\safemath{\matT}{\bT}
\safemath{\matU}{\bU}
\safemath{\matV}{\bV}
\safemath{\matW}{\bW}
\safemath{\matX}{\bX}
\safemath{\matY}{\bY}
\safemath{\matZ}{\bZ}
\safemath{\matZero}{\bZero}

\safemath{\matDelta}{\bDelta}
\safemath{\matLambda}{\bLambda}
\safemath{\matPhi}{\bPhi}
\safemath{\matSigma}{\bSigma}
\safemath{\matOmega}{\bOmega}
\safemath{\matTheta}{\bTheta}

\safemath{\matIdentity}{\matI}

\safemath{\complex}{\mathbb{C}}

\safemath{\timestep}{t}
\safemath{\femto}{s}
\safemath{\femtonumber}{S}
\safemath{\cohband}{B_c}
\safemath{\bandwidth}{B}

\safemath{\user}{k}
\safemath{\notuser}{\setminus\user}
\safemath{\other}{j}
\newcommand{\usernumber[1]}{K_{#1}}
\safemath{\userset}{\mathcal{\usernumber[]}}
\safemath{\carrier}{n}
\safemath{\othercarrier}{\ell}
\safemath{\notcarrier}{\setminus\carrier}
\safemath{\satcarrier}{\carrier_s}
\safemath{\actcarrier}{\carrier_a}
\safemath{\carriernumber}{N}
\safemath{\antenna}{m}
\safemath{\antennanumber}{M}
\safemath{\carrierset}{\mathcal{\carriernumber}}

\safemath{\curpower}{p_c}
\safemath{\noisepower}{\sigma^2}
\safemath{\sinrgap}{\Gamma}

\safemath{\powermatReq}{\mathbf{P}^{\textrm{req}}}

\safemath{\game}{\mathcal{G}}

\begin{document}
\IEEEoverridecommandlockouts
\title{{Distributed power control over interference channels using ACK/NACK feedback}}
\author{
\IEEEauthorblockN{Riccardo Andreotti\IEEEauthorrefmark{1},
Leonardo Marchetti\IEEEauthorrefmark{1}, Luca Sanguinetti\IEEEauthorrefmark{1}\IEEEauthorrefmark{2} and M{\'e}rouane Debbah\IEEEauthorrefmark{2}
\thanks{L.~Sanguinetti is funded by the People Programme (Marie Curie Actions) FP7 PIEF-GA-2012-330731 Dense4Green. This research has been supported by
the FP7 NEWCOM\# (Grant no. 318306), the ERC Starting MORE (Grant no.
305123), the French p\^ole de comp\'etitivit\'e SYSTEM@TIC within the
project 4G in Vitro and the project BESTCOM.}
}
\IEEEauthorblockA{\IEEEauthorrefmark{1}\small{Dipartimento di Ingegneria dell'Informazione, University of Pisa, Italy}}
\IEEEauthorblockA{\IEEEauthorrefmark{2}\small{Alcatel-Lucent Chair, Ecole sup{\'e}rieure d'{\'e}lectricit{\'e} (Sup{\'e}lec), Gif-sur-Yvette, France}}
}
\maketitle
%------------------------------------------------------------------------------------------------------------------
\begin{abstract}
In this work, we consider a network composed of several single-antenna transmitter-receiver pairs in which each pair aims at selfishly minimizing the power required to achieve a given signal-to-interference-plus-noise ratio. This is obtained modeling the transmitter-receiver pairs as rational agents that engage in a non-cooperative game. Capitalizing on the well-known results on the existence and structure of the generalized Nash equilibrium (GNE) point of the underlying game, a low complexity, iterative and distributed algorithm is derived to let each terminal reach the GNE using only a limited feedback in the form of link-layer acknowledgement (ACK) or negative acknowledgement (NACK). Numerical results are used to prove that the proposed solution is able to achieve convergence in a scalable and adaptive manner under different operating conditions. \end{abstract}

%{\keywords{Game theory, reinforcement learning, limited feedback, ACK, NACK, distributed algorithm}}

\section{Introduction}
The power consumption of the communication technology industry is becoming a major societal and economical concern \cite{EARTH_D23}, which has stimulated academia and industry to an intense activity in the new research area of green cellular networks \cite{Chen2011a,Smart_2020,GreenTouch}. The ultimate goal is to design new innovative network architectures and technologies needed to meet the explosive growth in cellular data demand without increasing the power consumption. Along this line of research, in this work we focus on a network composed of several single-antenna transmitter-receiver pairs operating over the same frequency band (or time slot) in which each pair aims at selfishly minimizing the power required to achieve a given signal-to-interference-plus-noise ratio (SINR). The mutual interference due to the simultaneous transmissions gives rise to a sort of competition for the common resource. The natural framework to study the solution of such interactions is non-cooperative game theory \cite{fud91,lasaulce-book-2011} in which the transmitter-receiver pairs are modeled as players that engage in a game using their own local information while fulfilling the given requirements. The existence and uniqueness properties of the equilibrium points of the underlying game have been widely studied in the literature and a large number of works already exist on this topic \cite{Mia10,Mia11, BBS14}. Particular attention has also been devoted to derive schemes based on best response dynamics that allow each player to achieve the equilibrium in an iterative and distributed manner. All these schemes rely on the assumption that the transmitter has perfect knowledge of the SINR measured at the receiver. This assumption does not hold true in practical applications and the only way for the transmitter to acquire this knowledge is through a return control channel. Although possible, however, this solution is not compliant with most of the current wireless communication standards in which the receiver only sends back a link-layer acknowledgement (ACK) whenever it is able to correctly decode the message and a negative ACK (NACK) otherwise. Most of the existing works dealing with resource allocation schemes using a 1-bit feedback are for centralized networks (see \cite{kok12,agg12}, and references therein), while only a few examples exist in decentralized scenarios. A first attempt in this direction is represented by \cite{huang10} where the authors propose a distributed power control
algorithm maximizing the sum rate in a secondary network,
under a given outage probability at the primary user. The latter
is evaluated by the secondary user by means of the $1$-bit
ACK/NACK feedback sent on the reverse link between the primary receiver
and transmitter. In \cite{bel10}, a distributed power
allocation scheme for outage probability minimization in multiple-input multiple-output (MIMO) interference
channels is proposed. The optimization problem is modeled as a non-cooperative
game with mixed strategies, where the probability of playing a
certain strategy is updated with a reinforcement learning rule
based on the ACK/NACK feedback.

The major contribution of this work is a novel iterative and distributed algorithm that allows the transmitters to converge to the equilibrium point using only the limited feedback in the form of ACK or NACK over packet-oriented transmission links. The proposed solution relies on a learning algorithm that is reminiscent of the scheme proposed in \cite{kok12} and allows each transmitter to locally update an estimate of the received SINR while converging towards the equilibrium. This is achieved applying a simple updating rule completely unaware of the structure of the underlying game and requiring knowledge of local information only. Numerical results are used to assess the convergence and performance (in terms of number of iterations required) of the proposed solution in the uplink of a small-cell network.

%It is worth mentioning that several works in the literature deal with link resource adaptation based on $1$-bit
%feedback in centralized scenarios. Examples in this field are given by \cite{kok12} and \cite{agg12},
%and references therein. Concerning distributed approaches,
%\cite{huang10} proposes a distributed power control
%algorithm that maximizes the sum rate in a secondary network,
%under a given outage probability at the primary user. The latter
%is evaluated by the secondary user by means of the $1$-bit
%ACK/NACK feedback sent on the reverse link between the primary receiver
%and the primary transmitter. In \cite{bel10}, a distributed power
%allocation for outage probability minimization in MIMO interference
%channels is proposed. The problem is modeled as a non-cooperative
%game with mixed strategies, where the probability of playing a
%certain strategy is updated with a reinforcement learning rule
%based on the ACK/NACK feedback.

\section{System model and problem formulation \label{sec:syst_model}}
We consider a $K-$user Gaussian interference channel, in which there are $K$ transmitter-receiver pairs sharing the same Gaussian channel, that might represent a time or frequency bin. The transmission is organized in frames with each frame counting a certain number of packets, each one composed of $M$ data symbols of unity-energy.
We call $x_{k}(m)$ the $m$th data symbol of transmitter $k$ within a generic packet and denote $\mathbf{x}_{k} = [x_{k}(1),x_{k}(2),\ldots,x_{k}(M)]^T$. Each $\mathbf{x}_{k}$ is encoded at a rate
$r_k \in \mathcal{R}_k$ with $\mathcal{R}_k$ being the set
of feasible rates and is transmitted with an amount of power $p_k \in \mathbb R_+$. The channel is assumed to be constant over a frame and to change independently from one frame to another (block-fading channel). We assume that the transmitters do not have any a-priori knowledge of the channel.

Letting $h_{k,i}$ denote the channel coefficient between transmitter $i$ and receiver $k$ over a generic packet, the vector $\mathbf{y}_{k} \in \mathbb{C}^{M\times 1}$ received at the $k$th receiver within the generic packet can be written as
\begin{equation}
\mathbf{y}_{k} = \sum\limits_{i=1}^K {h_{k,i}\sqrt {p_i}\mathbf{x}_{i}} + \mathbf{w}_{k}
\label{eq:received_signal}
\end{equation}
where $\mathbf{w}_{k} \sim \mathcal{CN} (0,{\sigma^2}\mathbf{I}_M)$ accounts for the additive white Gaussian noise. The corresponding SINR is given by
\begin{equation}
\gamma _k = \frac{p_k{|h_{k,k}{|^2}}}{{\sum\limits_{i=1, i \ne k}^{K} {|h_{k,i}{|^2}p_i + \sigma^2} }}.
\label{eq:sinr}
\end{equation}
For later convenience, we call
\begin{equation}
\mu _k = \frac{{|h_{k,k}{|^2}}}{{\sum\limits_{i=1, i \ne k}^{K} {|h_{k,i}{|^2}p_i + \sigma^2} }}
\label{eq:cinr}
\end{equation}
the channel-to-interference-plus-noise ratio (CINR) and denote $\mathbf{p}_{-k} = [p_1,\ldots,p_{k-1},p_{k+1},\ldots,p_{K}]^T$ the vector collecting all the transmit power except that of transmitter $k$. 

The aim of this work is to solve the following power minimization problem for any $k=1,2,\ldots,K$ 
\begin{align}\label{game}
\mathop {\min }\limits_{p_k \in \mathbb R_+} &  \quad p_k \\ \nonumber
{{\text{subject to}}} & \quad \gamma_k \ge \bar \gamma_k
\end{align}
where $\bar \gamma_k >0$ are given quality-of-service (QoS) requirements. The interplay among the pairs through \eqref{eq:sinr} makes \eqref{game} a multidimensional optimization problem in which each transmitter-receiver pair aims at unilaterally choosing the minimum transmit power $p_k$ so as to full fill its own requirement. In doing this, each pair affects the choice of all other pairs as well.

%{\bf{Remark 1}} The power consumption is a major issue in the uplink of wireless communications systems since user equipments (UEs) are usually
%low-powered and battery-limited devices. In small cell networks, where small cells are relatively close to each other, keeping the transmit power as low as possible has
%also the beneficial effect of limiting interference.

%While reducing the power consumption, however, we must take into account that UEs usually require certain quality of services (QoS) to be satisfied.
%
%Motivated by the above considerations, in this work we aim at solving the following power minimization problem
%\begin{align}\label{game}
%\mathop {\min }\limits_{p_k \in \mathbb R_+} &  \quad p_k \\ \nonumber
%{{\text{subject to}}} & \quad \gamma_k \ge \bar \gamma_k \quad k=1,2,\ldots,K
%\end{align}
%where $\bar \gamma_k >0$ for $k=1,2,\ldots,K$ are the given QoS requirements.

%
%By looking at the QoS constraint (\ref{eq:game}.b),
%we can note that the choice of the power of each user $q\in\calQ$
%affects the choice of all others as well,
%making this problem a multidimensional optimization problem.
%
%The above problem has been extensively studied in the literature,
%starting from the seminal work in \cite{yat95}. Most of the works rely on perfect channel knowledge at the UEs. For completeness, we recall
%its main features in the sequel. First of all, the natural framework to study this
%kind of interactions is that of

\section{Game Formulation\label{sec:game}}
%A close inspection of \eqref{game} reveals that it is a multidimensional optimization problem due to the coupling constraints.

The natural framework to study the solution of problems in the form of \eqref{game} is non-cooperative game theory \cite{fud91}. Interpreting \eqref{game} as a game \game leads to the definition of the tuple
$\game = (\calQ, \{\calA_k\}, \{u_k\})$, where $\calQ = \{1,2,\ldots,K\}$ is the set of players,
$\calA_k$ is the $k$th player's strategy set such that the constraints
in \eqref{game} are satisfied, and
$u_k=p_k$ is the utility function of player $k$.
Note that player $k$'s action set depends on the actions of the other players, i.e.,
$\calA_k=\calA_k({\bf p}_{-k})$ due to the presence of coupling constraints.
In this case, the solution concept to be used is the generalized Nash equilibrium (GNE)  that is defined as the point collecting all the system states stable to unilateral deviations \cite{fac07}.  The GNE of the power allocation problem in \eqref{game} has been extensively studied in the literature. The main results are summarized in the following theorem.

{\bf Theorem 1} \emph{If the problem \eqref{game}
is feasible, then there exists a unique power allocation vector ${\bf p}^*=[p_1^*,p_2^*,\ldots,p_K^*]^T$ that is the GNE of the game $\mathcal G$. The elements of ${\bf p}^*$ are the solutions to the following fixed-point system of
equations:
\begin{equation}
p_k^*={\rm BR}({\bf p}_{-k}^*)=\frac{\bar \gamma_k}{\mu_k({\bf p}^*_{-k})} \;\;\; \forall k \in \calQ
\label{eq:best_resp}
\end{equation}
where the operator ${\rm BR}$ stands for the best-response of user $k$
to given other users' strategy ${\bf p}_{-k}^*$ and $\mu_k$
is defined in (\ref{eq:cinr}).}

%The feasibility condition of \eqref{game} ensures that the overall set of
%strategies $\calA = \calA_1\times\cdots\times\calA_K$ is not empty.

As shown in \cite{pil05}, a necessary and sufficient condition for which problem
\eqref{game} is feasible is that $\rho_{\bf G}<1$ where $\rho_{\bf G}$ is the spectral radius of 
matrix ${\bf G}\in {\mathbb R}^{K\times K}$, whose $(k,i)$th element is computed as
\begin{equation}
{\left[ {\bf{G}} \right]_{k,i}} = \left\{ {\begin{array}{*{20}{c}}
0&{k = i}\\
{\frac{{\bar \gamma _k|{h_{k,i}}{|^2}}}{{|{h_{k,k}}{|^2}}}}&{k \ne i.}
\end{array}} \right.
\end{equation}
The existence and the uniqueness follows observing that the best-response is a standard function \cite{yat95}, \cite{sar02}.

%
%
%The feasibility of \eqref{game} amount to providing conditions under which the overall set of
%strategies $\calA = \calA_1\times\cdots\times\calA_K$
%is not empty. Upon defining the matrix
%${\bf G}\in {\mathbb R}^{K\times k}$, whose $(k,i)$th
%element is
%\begin{equation}
%{\left[ {\bf{G}} \right]_{k,i}} = \left\{ {\begin{array}{*{20}{c}}
%0&{k = i}\\
%{\frac{{\bar \gamma _k|{h_{k,i}}{|^2}}}{{|{h_{k,k}}{|^2}}}}&{k \ne i}
%\end{array}} \right.
%\end{equation}
%the necessary and sufficient condition for which problem
%\eqref{game} is feasible is that $\rho_{\bf G}<1$ \cite{pil05} with $\rho_{\bf G}$ being the spectral radius of ${\bf G}$.

In addition to this, using the results of \cite{yat95} it follows that the optimal point
${\bf p}^*$ can be reached via a distributed iterative
power control policy based on best response dynamics according to which
every player $k$ updates its power (strategy) $p_k^{(n+1)}$ at time $n+1$ as
\begin{equation} \label{power_update}
p_k^{(n+1)}=\frac{\bar \gamma_k}{\mu_k({\bf p}^{(n)}_{-k})}
\end{equation}
with $\mu_k({\bf p}^{(n)}_{-k})$ being the CINR within the transmission time $n$.

\section{Distributed algorithm \label{sec:algorithm}}
Using \eqref{eq:sinr}, we may rewrite \eqref{power_update} as
\begin{equation}
p_k^{(n+1)}=p_k^{(n)} \frac{\bar \gamma_k}{\gamma_k({\bf p}^{(n)})}
\end{equation}
from which it follows that the computation of $p_k^{(n+1)}$ for a given ${\bf p}^{(n)}_{-k}$ requires knowledge of $\gamma_k({\bf p}^{(n)})$. Most of the existing works rely on the assumption that each transmitter has perfect knowledge of it. Unfortunately, this assumption does not hold true in practical applications and the only way for the transmitter to acquire this knowledge is through a return control channel. Although possible, however, this solution is not compliant with current cellular standards in which the receiver only sends back an ACK $(f_k=0)$ whenever is able to correctly decode the packet and a NACK $(f_k=1)$ otherwise. Assume that a maximum likelihood (ML) decoder is used at the receiver and denote by ${{\bf{\hat x}}_k} \in \mathbb{C}^{M\times 1}$ the ML estimate of ${{\bf{x}}_k}$ obtained from
$\mathbf{y}_{k}$. Therefore, an ACK or NACK is sent to transmitter $k$ with probability
\begin{equation}
\Pr \left\{f_k = \bar f \right\} = \left\{ {\begin{array}{*{20}{l}}
{{\varepsilon _k}(\mu _k,r_k,{p}_k)}&{\bar f = 1}&{}\\
{1 - {\varepsilon _k}(\mu _k,r_k,{p}_k)}&{\bar f = 0}&{}
\end{array}} \right.
\end{equation}
where ${\varepsilon _k}(\mu _k,r_k,{p}_k)$ stands for the ML decoding error probability, which is clearly a function of the CINR $\mu _k$, the transmit power ${p}_k$ and the encoding rate $r_k$. In particular, assuming Gaussian
random codes a generic ${\varepsilon _k}(\mu _k,r_k,{p}_k)$ can be approximated as follows \cite{gal68}
\begin{equation}\nonumber
{{\varepsilon_k }(\mu _k,r_k,{p}_k)} \approx \exp\left( M\rho\left[ r_k\log 2 - \frac{1}{2}
\log\left(1+\frac{\gamma_k}{1+\rho}\right)
\right] \right)
\end{equation}
where $\rho\in[0,1]$ is the union bound parameter and $M$ is the number of data symbols per packet
encoded at a rate of $r_k$ bits/symbol.
% packet size $M$ as well as of the modulation and coding scheme adopted at the UE $k$.
%For the ease of notation, we call ${\bf I}_k^{(n)}= [{{\bf p}_k^{(n-1)}},{{\bf r}_k^{(n-1)}},{{\bf f}_k^{(n-1)}}]$ the matrix collecting all this information at transmission time of packet $n$.
%Next, we propose a two-step distributed procedure, which only relies on knowledge of $\{{{\bf p}_k^{(n-1)}},{{\bf r}_k^{(n-1)}},{{\bf f}_k^{(n-1)}}\}$.
%In the first step, each UE $k$ in $\calQ$ aims at locally computing a reliable estimate $\hat \mu^{(n)}_k$
%of $\mu^{(n)}_k$. This is then used in the second step for updating the transmit power according
%to \eqref{power_update}.

\begin{algorithm}[t]
\caption{Distributed resource allocation algorithm}
\begin{enumerate}
\item At $n=1$ for any $ k \in
\mathcal{K}$, choose a feasible $r_k^{(1)}\in\mathcal{R}_k$ and an arbitrary estimate $\hat\mu_k^{(1)}$. Then, set  $\hat p_k^{(1)} = {\bar \gamma_k}/{\hat\mu_k^{(1)}}$; \\
\item At $n=2,3, \ldots$ for any $k \in
\mathcal{K}$ \\
\begin{enumerate}
\item compute
\begin{align}\nonumber
{\hat \mu ^{(n)}_k} = {\hat \mu ^{(n - 1)}_k} + \frac{{{f_k^{(n - 1)}} -
\varepsilon ({{\hat \mu }_k^{(n - 1)}},{p_k^{(n - 1)}},{r_k^{(n - 1)}})}}{{{{(n - 1)}^\beta }
\varepsilon '({{\hat \mu }_k^{(n - 1)}},{p_k^{(n - 1)}},{r_k^{(n - 1)}})}}
\end{align}
and set
\begin{align}\nonumber
r_k^{(n)}=\mathop {\arg \max }\limits_{r \in {\mathcal R}_k} \Phi ({{\hat \mu }_k^{(n)}},{p_k^{(n-1)}},{r})
\end{align}
\vspace{0.05in}
\item update
\begin{align}\nonumber
p_k^{(n)}=\frac{\bar \gamma_k}{{{\hat \mu }_k^{(n)}}}
\end{align}
\end{enumerate}
\end{enumerate}
\end{algorithm}
%To proceed further, we denote ${\bf p}_k^{(n-1)}=[p_k^{(1)},p_k^{(2)},\ldots,p_k^{(n-1)}]^T$, ${\bf r}_k^{(n-1)}=[r_k^{(1)},r_k^{(2)},\ldots,r_k^{(n-1)}]^T$ and ${\bf f}_k^{(n-1)}=[f_k^{(1)},f_k^{(2)},\ldots,f_k^{(n-1)}]^T$ as the vectors collecting the previous values of powers, rates and feedbacks.
Based on the above considerations, we propose an iterative and distributed two-step algorithm that allows each transmitter-receiver pair to reach the GNE of the game only exploiting the knowledge of $\{{{p}_k^{(n-1)}},{{r}_k^{(n-1)}},{{f}_k^{(n-1)}}\}$. The first step is reminiscent of the iterative solution proposed in \cite{kok12} and
aims at locally computing a reliable estimate $\hat \mu^{(n)}_k$
of $\mu^{(n)}_k$. Mathematically, $\hat \mu^{(n)}_k$ is obtained as follows
\begin{align}\label{update_mu}
{\hat \mu ^{(n)}_k} = {\hat \mu ^{(n-1)}_k} + \frac{{{f_k^{(n - 1)}} -
\varepsilon ({\hat \mu ^{(n-1)}_k} ,{p_k^{(n - 1)}},{r_k^{(n - 1)}})}}{{{{(n - 1)}^\beta }
\varepsilon '({\hat \mu ^{(n-1)}_k} ,{p_k^{(n - 1)}},{r_k^{(n - 1)}})}}
\end{align}
where $\beta$ is a design parameter that regulates the convergence speed of the iterative procedure. The larger $\beta$, the smaller the convergence time. In addition, $\varepsilon '$ denotes the derivative of $\varepsilon$ with respect to $\mu$ and is given by
\begin{equation}\nonumber
\varepsilon ' = \frac{-M\rho p_k}{2(1+\rho+\mu_kp_k)}{{\varepsilon_k }}
\end{equation}
where we have dropped the functional dependence from $(\mu _k,r_k,{p}_k)$ for notational simplicity.
The value of ${\hat \mu ^{(n)}_k}$ is then used to update $r_k^{(n)}$ according to:
\begin{align}
r_k^{(n)}=\mathop {\arg \max }\limits_{r \in {\mathcal R}_k} \Phi ({{\hat \mu }_k^{(n)}},{p_k^{(n-1)}},{r})
\end{align}
where
\begin{equation}\nonumber
\Phi ({{\hat \mu }_k^{(n)}},{p_k^{(n-1)}},{r}) = \frac{{{{[\varepsilon '({{\hat \mu }_k^{(n)}},{p_k^{(n-1)}},{r})]}^2}}}{{\varepsilon ({{\hat \mu }_k^{(n)}},{p_k^{(n-1)}},{r})[1 - \varepsilon ({{\hat \mu }_k^{(n)}},{p_k^{(n-1)}},{r})]}}
\end{equation}
is the Fisher information associated to the random variable $f_k$. Following the same arguments of \cite{kok12}, it can be proven that for any unbiased estimator based on $n$ ACK/NACK, the estimation error variance of ${\mu ^{(n)}_k}$ is lower bounded by the reciprocal of the cumulative Fisher information given by $\sum\nolimits_{i=1}^n\Phi ({{ \mu }_k^{(i)}},{p_k^{(i-1)}},{r}_k^{(i)})$.

The estimate ${\hat \mu ^{(n)}_k}$ is eventually used in the last step for updating the transmit power as specified in \eqref{power_update}. The main steps of the proposed solution are summarized in {\bf{Algorithm 1}} where $f_k^{(1)}$ is the ACK/NACK received after the first packet transmission.

{\bf{Remark 1}}.
\emph{Observe that similarly to a reinforcement learning approach in which at each step
the probability function is updated according to a certain rule
and then a strategy is randomly played according to this
probability, in the proposed solution the estimate of ${\mu_k}$ is updated through \eqref{update_mu}, but
then, the strategy is deterministically played, exploiting
the knowledge of the optimal solution of the game with
complete information.}

{\bf{Remark 2}}. \emph{The analytical study of the convergence of
the proposed algorithm is still much open and left for future work. In the next section, we limit to assess the convergence of {\bf Algorithm 1} by means of Monte Carlo simulations. Interestingly, it turns out that the proposed solution converges (within the
required accuracy) whenever the game with complete information is feasible
and thus the existence of the unique GNE point is guaranteed. Moreover, the convergence point is the same meaning that the same performance can be achieved despite the amount of required information is much lower. The only price to pay is a greater convergence time.}

\section{Simulation Results \label{sec:sim_res}}

The performance of the distributed algorithm is now assessed by means of an extensive simulation campaign. To this end, we consider the uplink of a small-cell network \cite{hoy11} consisting of either $K= 4$ or $K=6$ single-antenna small cells, each serving a single user. We set $\boldsymbol{\bar \gamma} = [0.5, 1, 1.5, 2, 2.5, 3]^T$ dB and assume that the coverage area of each small cell is circular with radius $R=50$ m and minimum distance $R_{\min} = 5$ m. The small cells are randomly distributed over a $200\times50K$ area. Moreover, we consider a system in which the large-scale fading is dominated by the path-loss. This amounts to saying that the channel coefficients $h_{k,i}$ can be modeled as
\begin{equation}
h_{k,i} = \frac{\bar d}{d ^{\alpha}}{\bar h}_{k,i} \quad \text{for} \; d\geq R_{\min}
\end{equation}
where ${\bar h}_{k,i}\sim \mathcal {CN} (0,1)$ accounts for the small-scale fading, $d$ is the distance between transmitter $i$ and receiver $k$, $\alpha \geq 2$ is the path-loss exponent and $\bar d>0$ is a constant that regulates the channel attenuation at distance $R_{\min}$ \cite{LTE2010b}. We set $\bar{d}=10^{-3.53}$ and $\alpha=3.76$. We assume that the channel coefficients maintain constant in time. Moreover, the noise power level is set to $\sigma^2 = -100$ dBm and each packet is assumed to contain $M=500$ symbols. The proposed algorithm is initialized for any $k \in \mathcal K$ as follows: $r_k^{(1)} = 1 $ bit/s/Hz, $\hat\mu_k^{(1)} = |h_{k,k}|^2/\sigma^2$ and $p_k^{(1)} = \bar \gamma_k/\hat\mu_k^{(1)}$.

\begin{figure}[!t]
  \begin{center}
      \includegraphics[width=0.9\columnwidth]{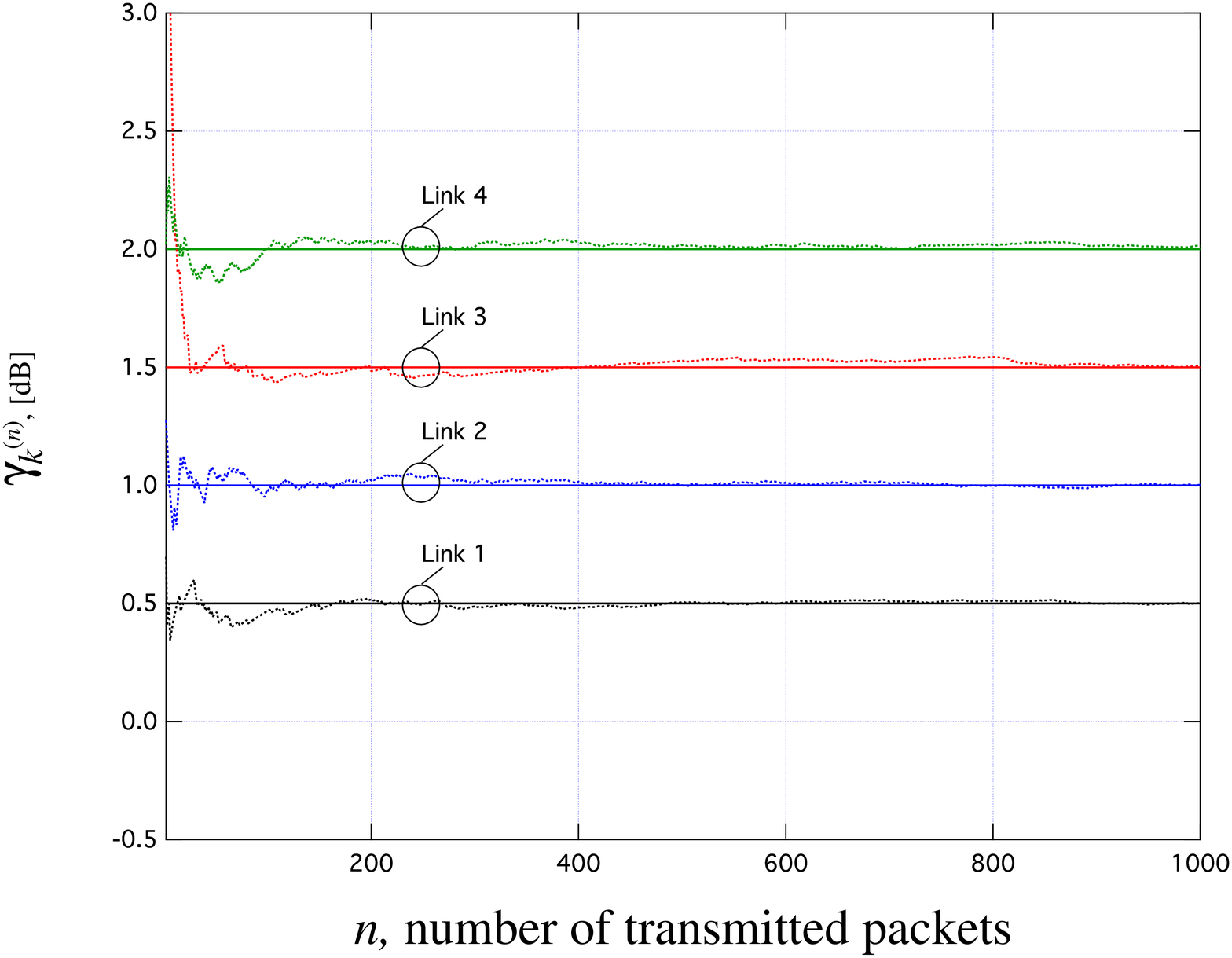}
  \caption{SINR vs. number of packets when $K=4$ and $\beta=0.9$.}
  \label{fig1}
  \end{center}
\end{figure}

\begin{figure}[!t]
  \begin{center}
    \includegraphics[width=0.9\columnwidth]{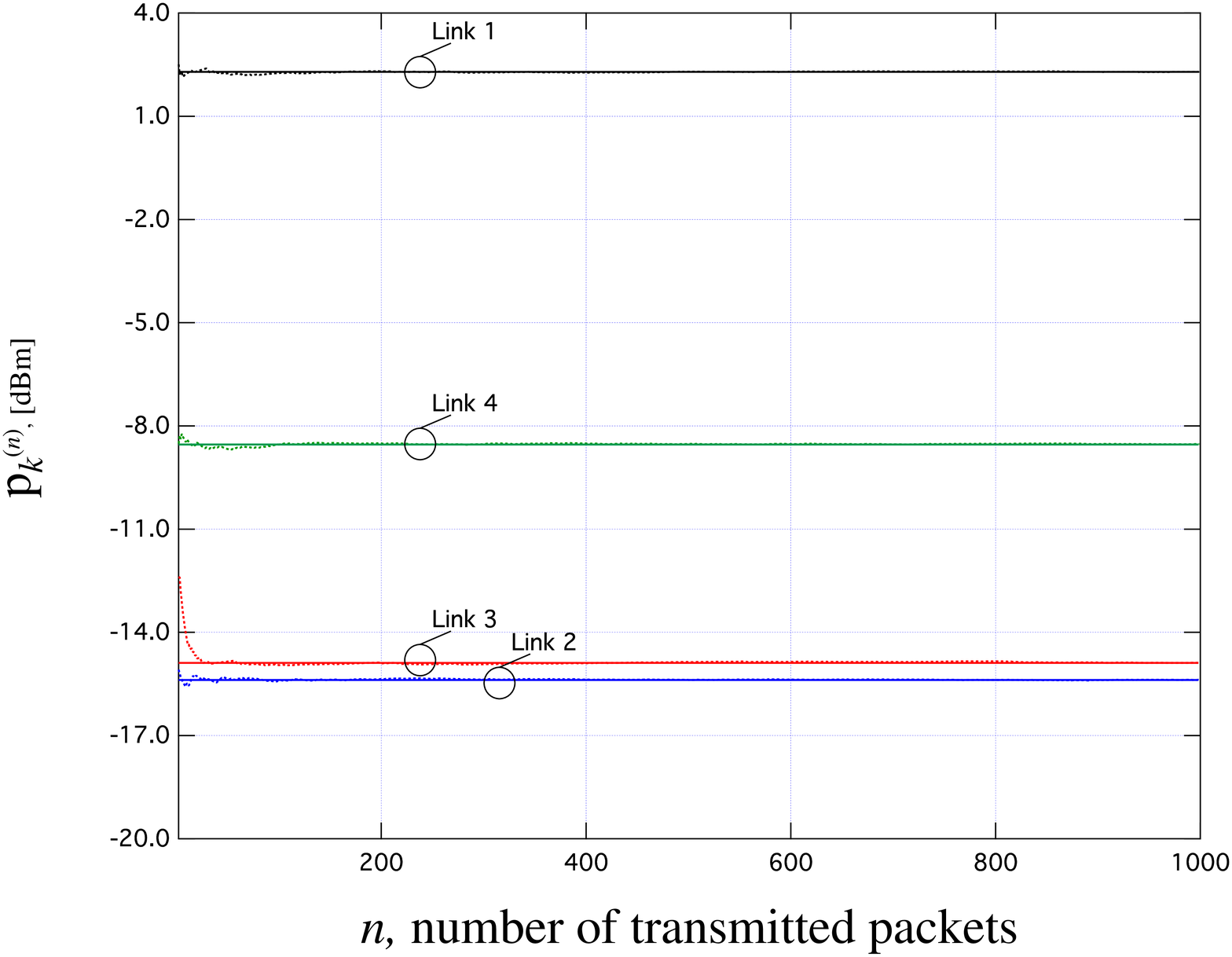}
  \caption{Transmit power vs. number of packets when $K=4$ and $\beta=0.9$.}
  \label{fig2}
  \end{center}
\end{figure}

\begin{figure}[!t]
  \begin{center}
    \includegraphics[width=0.9\columnwidth]{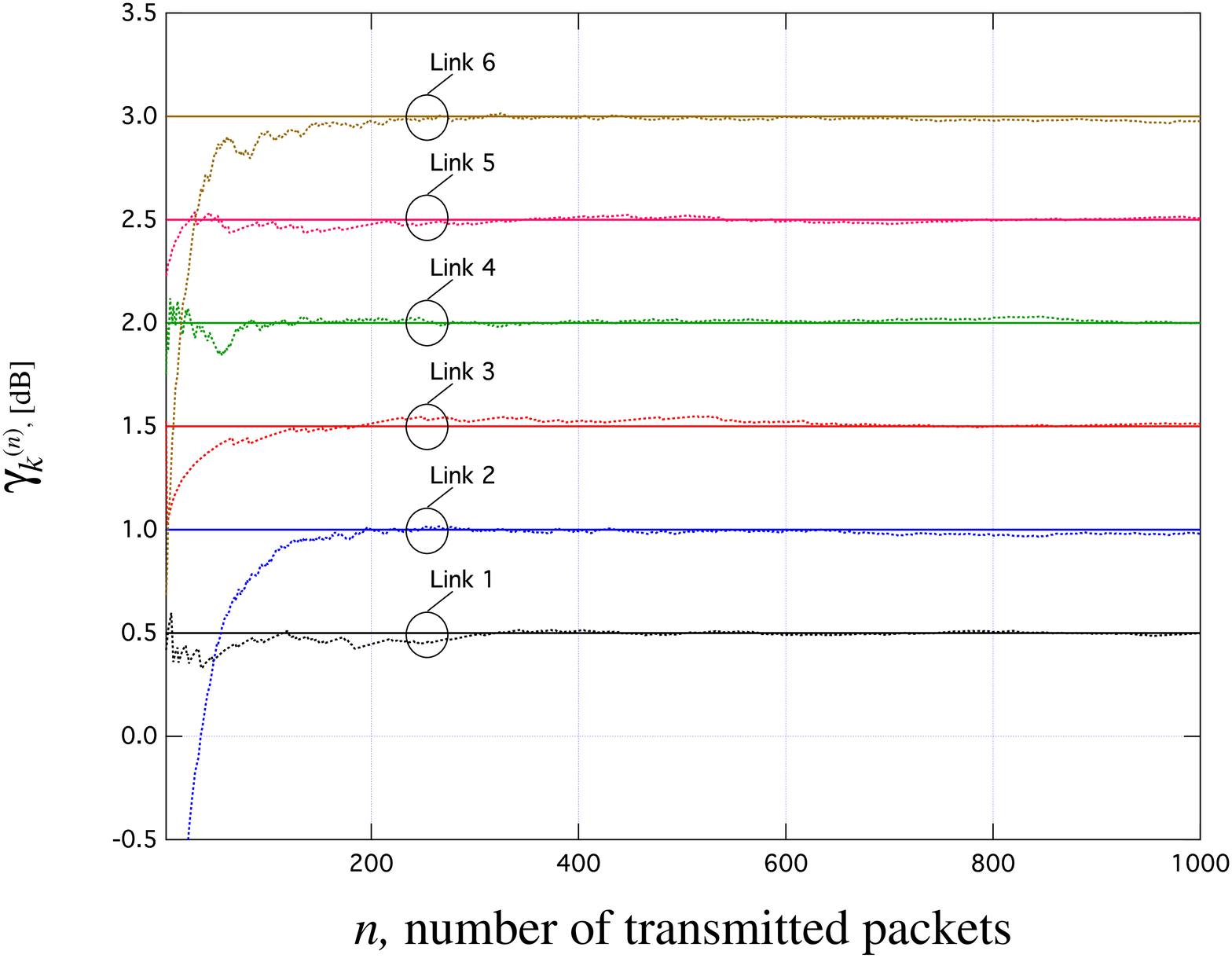}
  \caption{SINR vs. number of packets when $K=6$ and $\beta=0.9$.}
  \label{fig3}
  \end{center}
\end{figure}

\begin{figure}[!t]
  \begin{center}
    \includegraphics[width=0.9\columnwidth]{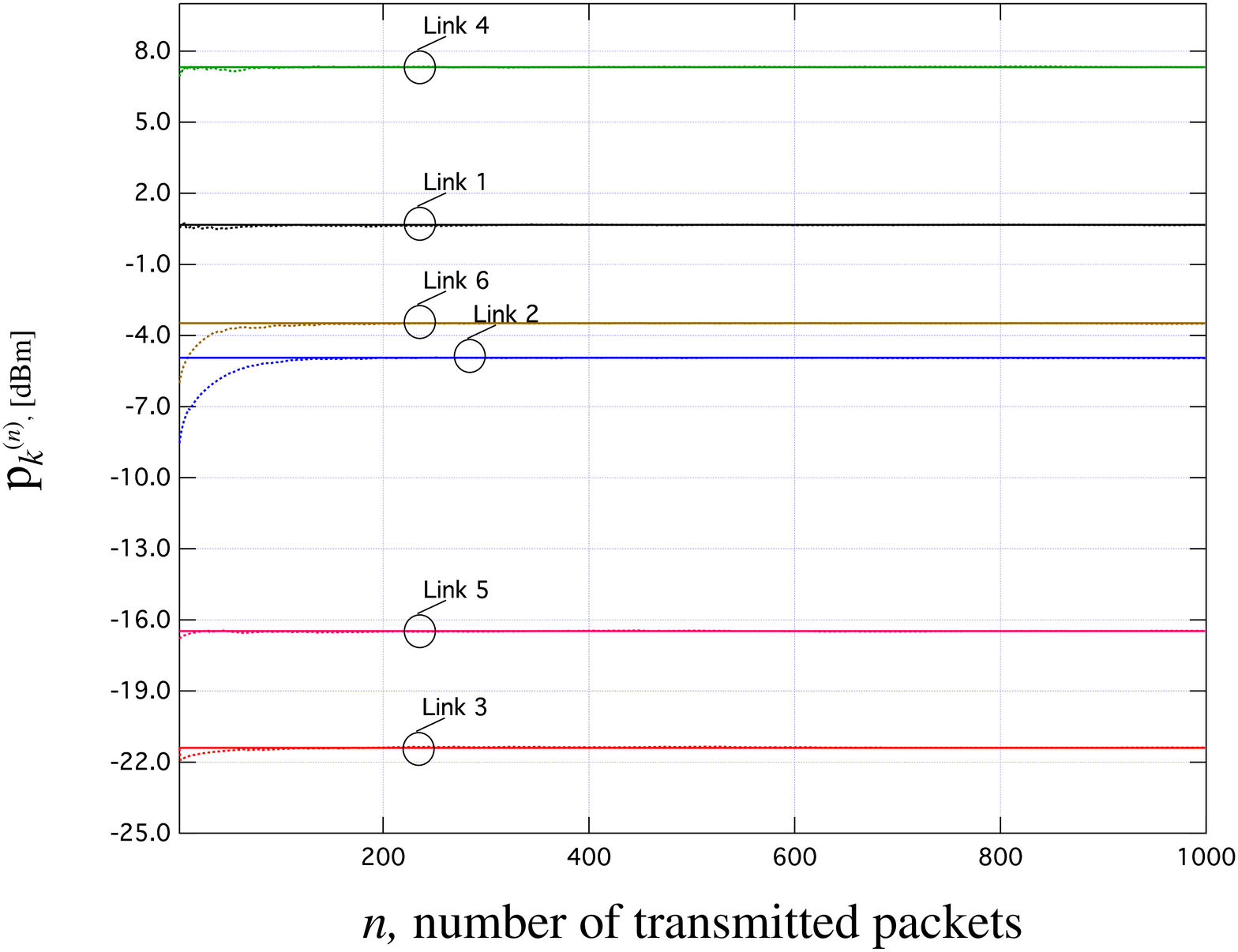}
  \caption{Transmit power vs. number of packets when $K=6$ and $\beta=0.9$.}
  \label{fig4}
  \end{center}
\end{figure}

Fig. \ref{fig1} illustrates the values of $ \gamma_k^{(n)}$ (dashed lines) measured at the small cell access point as a function of the number $n$ of transmitted packets in a scenario of $K=4$ small cells when $\beta =0.9$. The target SINRs $\bar \gamma_k$ for $k=1,2,\ldots,K$ (continuous lines) are also reported for comparison. Fig. \ref{fig2} reports also the variations of $ p_k^{(n)}$ (dashed lines) as $n$ increases together with the power (continuous lines) required at the GNE point. The results of Figs. \ref{fig3} and \ref{fig4} are obtained in the same operating conditions of Figs. \ref{fig1} and \ref{fig2} except that now $K=6$. As seen, in both cases, $\gamma_k^{(n)}$ converges to the target SINR $\bar \gamma_k$ within $200$ packets. Interestingly, the attained power level is exactly the same achieved at the GNE point of the game with complete information.

\begin{figure}[!t]
  \begin{center}
    \includegraphics[width=0.9\columnwidth]{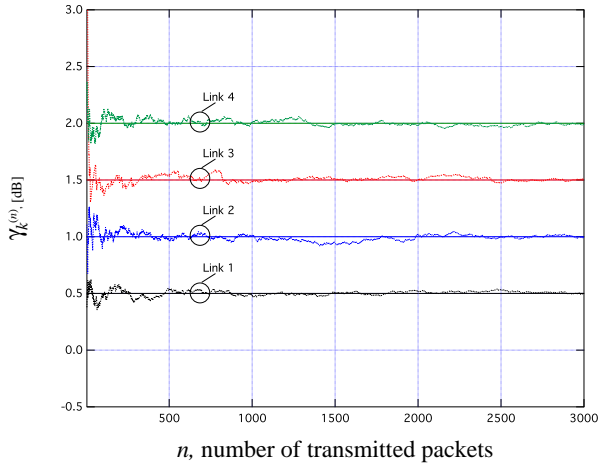}
  \caption{SINR vs. number of packets when $K=4$ and $\beta=0.5$.}
  \label{fig5}
  \end{center}
\end{figure}

The results of Fig. \ref{fig5} illustrates the behaviour of $\gamma_k^{(n)}$ when $K=4$ and $\beta$ is set to $0.5$. As expected, reducing $\beta$ allows terminals to achieve convergence within a smaller number of packets. However, this is achieved at the price of larger variations around the target values $\{\bar\gamma_k\}$.

\begin{figure}[!t]
  \begin{center}
    \includegraphics[width=0.9\columnwidth]{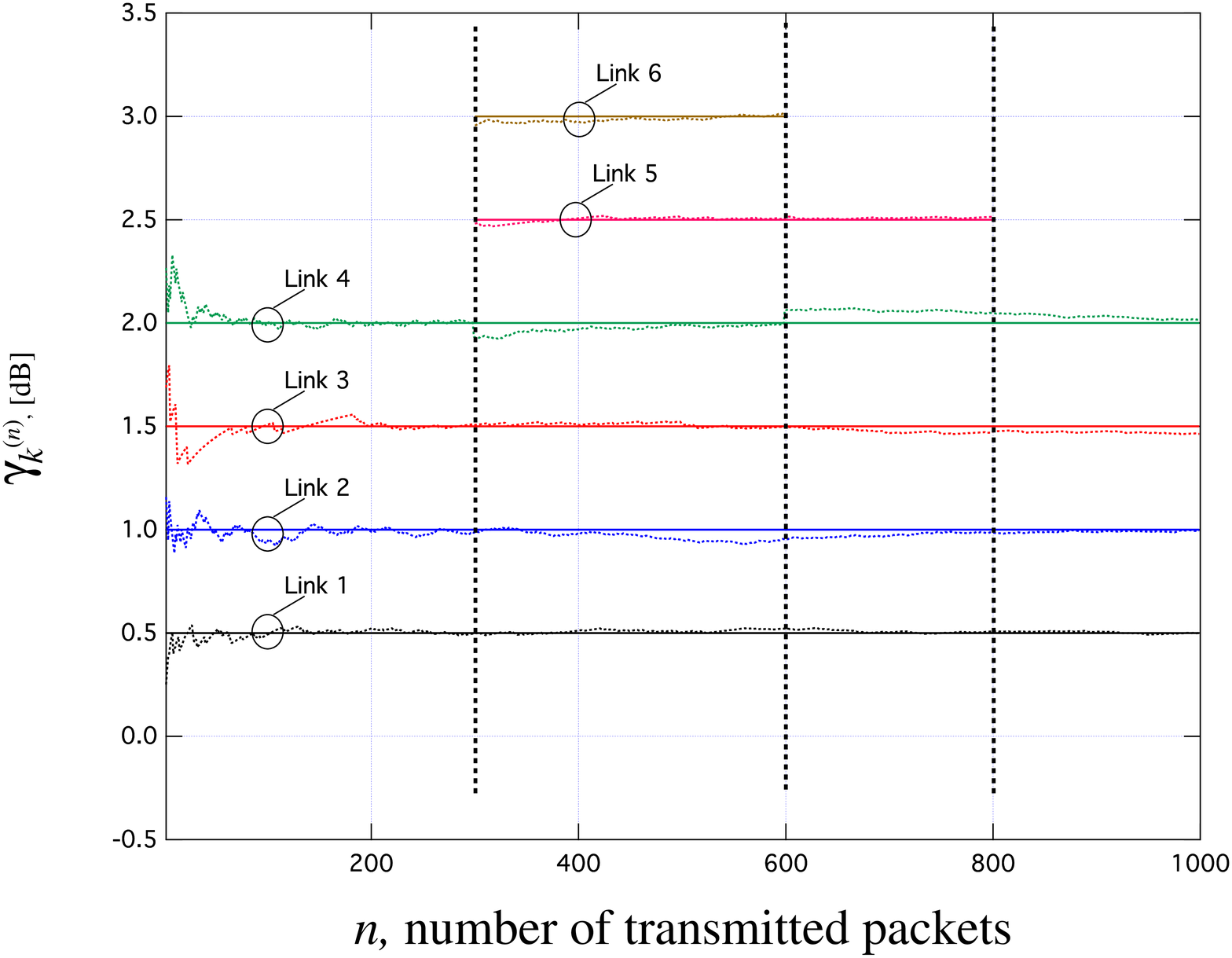}
  \caption{SINR vs. number of packets with $\beta=0.9$.}
  \label{fig7}
  \end{center}
\end{figure}

Fig. \ref{fig7} shows $ \gamma_k^{(n)}$ when $K$ changes during the execution of the algorithm. We start with $K=4$ at $n=1$ and assume that two new small cells become active at $n=300$ and inactive again at $n=600$ and $n=800$. As seen, the algorithm is very robust to network perturbations and guarantees fast convergence in a dynamic scenario.

\section{Conclusions \label{sec:concl}}

In this work, we have focused on the problem of selfishly minimizing the power consumption while satisfying target SINR constraints in interference channels characterized by single-antenna transmitter-receiver pairs operating over the same frequency band or time slot. In particular, we have first modeled the problem as a non-cooperative game with perfect channel state information and then we have solved it assuming that each transmitter has no knowledge about the propagation channel but could only exploit the ACK or NACK feedbacks generated at the link layer from the receiver. This choice has been motivated by the fact that it is compliant with many wireless communication standards and avoids the need of introducing a suitably designed return control channel. Accordingly, we have proposed an iterative and distributed algorithm inspired by best response dynamics in which (at each step) every transmitter updates its power exploiting a local estimate of its current SINR at the receiver. The latter is learned step by step via an updating rule based on the 1-bit feedback information given by ACK or NACK. The performance of the proposed solution have been evaluated by means of numerical results in the uplink of a small cell network. It turns out that the algorithm converges reasonably fast to the GNE point of the underlying game with perfect CSI. Further research is needed to provide an analytical proof about the convergence of the iterative procedure.

%%------------------------------------------------------------------------------------------------------------------
%\section*{Acknowledgment}
%This work was supported by the European Commission in the framework
%of the FP7 Network of Excellence in Wireless COMmunications NEWCOM\#
%(Grant agreement no. 318306).
%L. Sanguinetti has received funding from the People Programme (Marie Curie Actions) of the European Unions FP7 under REA Grant agreements no. PIEF-GA-2012-330731 ``Dense4Green''

%------------------------------------------------------------------------------------------------------------------

\end{document}